\documentclass[12pt,a4paper,final]{iopart}
  
\usepackage{iopams}  
\usepackage{graphicx}

\usepackage[breaklinks=true,colorlinks=true,linkcolor=blue,urlcolor=blue,citecolor=blue]{hyperref}
\usepackage[export]{adjustbox}
\usepackage{hyperref}
\usepackage{graphicx}
\usepackage{color}

\newcommand{\mbf}[1]{\mathbf{#1}}

\newcommand{\tmax}{t_{\textrm{max}}}

\begin{document}

\title{Quasi-one-dimensional Hall physics in the Harper-Hofstadter-Mott model}

\author{Filip Kozarski, Dario H\"ugel, and Lode Pollet}

\address{Department of Physics, Arnold Sommerfeld Center for Theoretical Physics and Center for NanoScience, Ludwig-Maximilians-Universit\"at M\"unchen, Theresienstrasse 37, 80333 Munich, Germany} 

\date{\today} 
\begin{abstract}

We study the ground-state phase diagram of the strongly interacting Harper-Hofstadter-Mott model at quarter flux on a quasi-one-dimensional lattice consisting of a single magnetic flux quantum in $y$-direction.
In addition to superfluid phases with various density patterns, the ground-state phase diagram features quasi-one-dimensional analogues of fractional quantum Hall phases at fillings $\nu=1/2$ and $3/2$, 
where the latter is only found thanks to the hopping anisotropy and the quasi-one-dimensional geometry. At integer fillings -- where in the full two-dimensional system the ground-state is expected to be gapless -- we observe gapped non-degenerate ground-states: 
At $\nu=1$ it shows an odd ``fermionic'' Hall conductance, while the Hall response at $\nu=2$ consists of the transverse transport of a single particle-hole pair, resulting in a net zero Hall conductance. The results
are obtained by exact diagonalization and in the reciprocal mean-field approximation. 

\end{abstract}

\maketitle
\makeatletter
\let\toc@pre\relax
\let\toc@post\relax
\makeatother

\section{Introduction}

The prospect of realizing and measuring topologically non-trivial bosonic phases remains an intriguing and important challenge of condensed matter physics. The bosonic statistics can lead to different topological properties than the ones observed in fermionic systems \cite{Cooper_PRL_01,ED_05,Cooper_Adv_08,Mol_CFM,XiaoGangWen_SPT_Science,Vish}. Furthermore, as bosons condense in the absence of interactions, topologically non-trivial bosonic phases are inherently many-body in nature, as an interaction is needed to introduce a gap. From an experimental point of view, bosonic atoms are easier to control in cold atom experiments with optical lattices \cite{Madi_ArtGa,Abo_Art,BlochRev,Lin_art,Dali_Art}, making them prime candidates in the search for interacting topological properties.

One promising lattice model -- which has already been experimentally realized in the non-interacting case \cite{Hof,Miyake_13,Atal_14,Aid_15} -- is the Harper-Hofstadter-Mott model (HHMm), the locally interacting version of the Harper-Hofstadter model \cite{Harper_HHM,Hof_HHM}. In this system previous works have predicted topologically non-trivial phases both of a long-range entangled intrinsic nature -- such as the fractional quantum Hall (fQH) phase at filling (per unit-cell) $\nu=1/2$ observed with exact diagonalization (ED) \cite{ED_05,Mol_CFM,ED_07} and the density matrix renormalization group (DMRG) \cite{He_HHM,dmrg_hhm2,dmrg_hhm1} -- as well as of short-range entangled symmetry protected kind \cite{XiaoGangWen_SPT_Science,Vish}, such as the recently predicted bosonic integer quantum Hall phase at filling $\nu=2$ \cite{He_HHM}.

While both ED and DMRG have provided great insight in such models, they rely on finite system sizes. In two dimensions, these may be too small for ED whereas DMRG has difficulties converging for gapless phases due to rapid entanglement growth; i.e., there is a preference for gapped, low entanglement phases. In order to obtain the full phase diagram, we see therefore a need to develop new methods, such as the recently proposed reciprocal cluster mean-field (RCMF) method, which has also been applied to the two-dimensional HHMm \cite{Hugel17}, and which tends to favor the opposite: it is defined in the thermodynamic limit and favors condensed phases (although it certainly can find topologically non-trivial phases as we will see). A systematic comparison of the two approaches therefore provides a promising path towards understanding the ground-state properties of the system in question. Furthermore, the restriction of the $y$-direction to just a few sites (in our case four) enables us to benchmark our RCMF results against ED, where it suffices to scale the system-size in the $x$-direction only.

In this work we analyze the properties of the HHMm on a quasi-one-dimensional lattice, consisting of just a single flux quantum along the $y$-direction, while the $x$-direction is treated in the thermodynamical limit. For the flux of $\Phi=\pi/2$ considered here, this consists of $4$ plaquettes in $y$-direction with periodic boundaries. Such a thin-torus limit has been previously investigated in fermionic systems in the lowest Landau level, where one-dimensional analogues of quantum Hall states were observed, which are predicted to continuously develop into their two-dimensional counterparts for increasing $y$-direction \cite{TT1,TT2,TT3}. For interacting bosons an effective ladder model realizing the thin-torus limit with $2$ sites in $y$-direction has been proposed, predicting a charge density wave analogue of the two-dimensional $\nu=1/2$ fQH phase \cite{Grusdt_Lad}.

The quasi-one-dimensional limit in combination with anisotropic hopping amplitudes leads to larger many-body gaps due to the finite size in $y$-direction, and therefore to more stable topological phases than in in the fully two-dimensional limit. This can be a useful insight in the experimental search for bosonic topologically non-trivial phases, where robustness against the expected strong heating processes is of great importance~\cite{Bilitewski14}. 
Equally important, fillings which are expected to be always gapless in the fully two-dimensional limit \cite{Senthil,Lu_17} can become gapped as a consequence of the finite size, leading to unexpected new ground-states. 
Another feature of the quasi-one-dimensional geometry lies in its low number of sites in $y$-direction, which makes it possible to map the spatial $y$-direction onto a finite number  of internal degrees of freedom (in this case 4), rewriting the system as a one-dimensional multi-component system, which in the future could be simulated by cold atoms in the  synthetic dimensions concept~\cite{Boada2012,Celi2014, Mancini2015,Stuhl2015}, or by using microwave cavities~\cite{JonSimon2016}. 

While at other fillings our ground-state phase diagram features superfluid phases with striped or checkerboard order, at integer and half filling we observe a number of gapped phases in agreement with ED results. At $\nu=1/2$ and $3/2$ we find that the quasi-one-dimensional geometry in combination with hopping anisotropy stabilizes gapped degenerate ground-states, which are quasi-one-dimensional analogues of fQH phases with a quantized fractional Hall response, differing from their two-dimensional counterparts in the continuum through a weak charge density wave order. At integer fillings -- where for the flux considered here the uncondensed phases are always gapless in the two-dimensional setup \cite{Hugel17} -- we see that the anisotropic setup introduces new phases with surprising properties: We observe gapped non-degenerate ground-states, which at $\nu=1$ feature a ``fermionic'' Hall conductance of $\sigma_{xy}=1$, while at $\nu=2$ the Hall response consists of the quantized transport of a single particle-hole pair with total conductance of $\sigma_{xy}=0$.

This paper is organized as follows. The HHMm on a cylinder is introduced in Sec.~\ref{sec:model}, while RCMF is quickly reviewed in Sec.~\ref{sec:rcmf}. In Sec.~\ref{sec:results} we present our results on the ground-state phase diagram and discuss the individual phases, providing a conclusion in Sec.~\ref{sec:conc}.

\section{Harper-Hofstadter-Mott model on a cylinder}\label{sec:model}

The Hamiltonian of the HHMm can be written as
\begin{eqnarray}\label{eq:HHMm}
H =& -\sum_{x,y} ( t_x e^{i y \Phi} b_{x+1,y}^\dagger b_{x,y} + t_y b_{x,y+1}^\dagger b_{x,y}) + {\rm H.c.}\\
& + \frac{U}{2} \sum_{x,y} n_{x,y} \left(n_{x,y}-1\right) - \mu \sum_{x,y} n_{x,y},
\end{eqnarray}
where the coordinates $(x,y)$ parameterize a system of size $L_x \times L_y$, with periodic boundary conditions. The operators $b_{x,y}^{(\dagger)}$ are the annihilation (creation) operators at site $(x,y)$, and $n_{x,y}=b_{x,y}^{\dagger}b_{x,y}$ is the occupation number operator. The hopping amplitudes in $x$- and $y$-direction are $t_x$ and $t_y$, respectively, $\Phi$ is the flux through each plaquette, $U$ is the strength of the on-site interaction, and $\mu$ is the chemical potential. 
Throughout our analysis we focus on the hard-core boson limit, $U\to\infty$, and the flux $\Phi=\pi/2$. We are left with two dimensionless parameters, the hopping anisotropy $\left(t_x-t_y\right)/t_{\rm max}$, and the chemical potential $\mu/t_{\rm max}$, where $\tmax=\max\{t_x,t_y\}$.

Furthermore, we focus on the cylinder-geometry in the thermodynamic limit, with $L_x \to\infty$ and $L_y=4$. In the case $\Phi=\pi/2$ the magnetic unit-cell is of size $1\times4$ for the Landau gauge used in Eq.\ (\ref{eq:HHMm}). This choice of the lattice size therefore makes the cylinder quasi-one-dimensional, as only one unit-cell is present in the $y$-direction, and different ground-state phases than those of the fully two-dimensional HHMm \cite{ED_05,Mol_CFM,ED_07,He_HHM,dmrg_hhm2,Hugel17,Palm_08,Onur1,Onur2} can be expected. Note that even though the magnetic unit cell can be chosen to consist of less than $4$ sites in $y$-direction, e.g.\ of size $2\times 2$ \cite{Mol_C2}, the particles would still need to perform a (gauge-invariant) loop around $4$ plaquettes in $y$-direction in order to pick up a trivial phase of $2\pi$. Having just two plaquettes in $y$-direction on the other hand, would correspond to just half of the flux quantum $2\pi$. In this sense our approach to quasi-one-dimensionality is different to the one of Refs.\  \cite{Grusdt_Lad,Lad_Belen,Lad_Marie}, where there is just one plaquette in $y$-direction on a ladder geometry.

The Hamiltonian (\ref{eq:HHMm}) can be block-diagonalized in the hard-core boson limit with the Fourier transform $d_y (k) = \sqrt{L_x}^{-1}\sum_{x} e^{-i k x} d_{x,y}$, where $k \in \big\{\frac{2 \pi m}{L_x} \bigm| m \in \{0, \dots , L_x -1 \}\big\}$ is the quasi-momentum in the $x$-direction, and $d_{x,y}$ is the annihilation operator of a hard-core boson at site $(x,y)$ with $\left(d_{x,y}\right)^{2}=\left(d_{x,y}^{\dagger}\right)^{2}=0$. In this basis Eq.\ (\ref{eq:HHMm}) for the case $\Phi=\pi/2$ can be rewritten as $H = \sum_{k} H_k$, where the Harper Hamiltonian \cite{Harper_HHM} $H_k$ is given by
\begin{equation}\label{eq:Ham-diag}
H_k = -\sum_{y=0}^3 \left(t_x e^{i (\frac{\pi}{2} y -k)} d_y^\dagger(k) d_y(k)+t_y  d_{y+1}^\dagger(k) d_y(k)\right) +{\rm H.c.}
\end{equation}
This Hamiltonian can be further written in terms of the $h$-vector \cite{Hugel17}, which measures the Hall response -- see  ~\ref{sec:twisted} for details.

Note that the Hamiltonian (\ref{eq:HHMm}) in the hard-core boson case is invariant -- up to a constant -- under the charge-conjugation transformation
\begin{equation}\label{eq:ct}
d_{x,y} \leftrightarrow d^\dagger_{x,y},\quad \Phi \mapsto -\Phi,\quad \mu \mapsto -\mu.
\end{equation}
This implies that the ground-states at positive and negative chemical potentials (or equivalently at densities $n$ and $1-n$), are related by the holes taking on the role of particles and the Hall response changing sign ($\sigma_{xy}\mapsto-\sigma_{xy}$) \cite{Hugel17,Lind_10}. In the following we will therefore restrict ourselves to the case of $\mu\leq 0$ (i.e.\ $n\leq 1/2$), with the phases at positive chemical potentials related to the ones discussed in this work by Eq.\ (\ref{eq:ct}).

\section{Reciprocal Cluster Mean-Field Theory}\label{sec:rcmf}

The main method we employ for the analysis of the model is RCMF \cite{Hugel17}, whose results we will further support by ED. It is defined in the thermodynamic limit and variationally approaches both condensed and uncondensed phases in models with multiorbital unit-cells and non-trivial dispersions, while preserving the translational symmetry of the lattice. Within this method, a Hamiltonian such as (\ref{eq:HHMm}) in the thermodynamic limit is decoupled into a set of identical clusters with size $C_x\times C_y$ (here $C_x=C_y=4$) through a combination of momentum coarse-graining \cite{Quant_Clust} and the mean-field decoupling approximation \cite{MF_Fisher}.

The method can be applied to Hamiltonians of the form
\begin{equation}\label{eq:Hgeneral}
H=\sum_{x,x'} \sum_{y,y'} t_{(x,y),(x',y')}\; b^{\dagger}_{x,y} b_{x',y'} + H_{\rm{int}},
\end{equation}
with some local interaction term $H_{\rm{int}}$ and translational-invariant (up to the unit cell) hopping amplitudes $t_{(x,y),(x',y')}$. Assuming the unit cell consists of $N_{\Phi}$ sites, the Hamiltonian in momentum space reads [see e.g.\ Eq.\ (\ref{eq:Ham-diag-Ly8})]
\begin{equation}\label{eq:Hepsilon}
H=\sum_{k,q} \sum_{\ell,\ell'} \varepsilon_{\ell,\ell'}(k,q)\; b^{\dagger}_{\ell}(k,q) b_{\ell'}(k,q) + H_{\rm{int}} ,
\end{equation}
where $\ell \in [0,N_{\Phi}-1]$ describes the position within the unit cell, $\varepsilon_{\ell,\ell'}(k,q)$ is a generalized $N_{\Phi}\times N_{\Phi}$ dispersion of the unit cell and $k$, $q$ are momenta in $x$, $y$ direction, respectively. We start by dividing the lattice of size $L_x \times L_y$ into $\frac{L_x L_y}{C_x C_y}$ clusters of size $C_x \times C_y$. As is done also in the dynamical cluster approximation  \cite{Quant_Clust}, we decompose the position coordinates as $(x,y) = (X+\tilde{x}, Y+\tilde{y})$ and the quasi-momenta as $(k,q)=(K+\tilde{k},Q+\tilde{q})$, where $(X,Y)$, $(K,Q)$ are the intra-cluster components, and $(\tilde{x},\tilde{y})$, $(\tilde{k},\tilde{q})$ are the inter-cluster components. The dispersion can now be rewritten as $\varepsilon_{\ell,\ell'}(K+\tilde{k},Q+\tilde{q}) = \overline{\varepsilon}_{\ell,\ell'}(K,Q) + \delta\varepsilon_{\ell,\ell'}(K+\tilde{k},Q+\tilde{q})$, where
\begin{equation}\label{eq:cluster-disp}
\overline{\varepsilon}_{\ell, \ell'}(K,Q)=\frac{C_x C_y}{L_x L_y}\sum_{\tilde{k},\tilde{q}} \varepsilon_{\ell,\ell'}(K+\tilde{k},Q+\tilde{q}) ,
\end{equation}
is the effective intra-cluster dispersion with periodic boundaries on the clusters.

The kinetic part of the Hamiltonian (\ref{eq:Hepsilon}) now consists of two terms, namely the cluster-local term with hopping amplitudes arising from $\overline{\varepsilon}_{\ell,\ell'}(K,Q)$, and the inter-cluster term capturing hopping processes between different clusters with the remainder of the dispersion $\delta\varepsilon_{\ell,\ell'}(K+\tilde{k},Q+\tilde{q})$.

In the next step we perform the \emph{mean-field decoupling} in the inter-cluster term, expanding the bosonic operators as $b_\ell(K+\tilde{k},Q+\tilde{q})=\phi_\ell(K+\tilde{k},Q+\tilde{q})+\delta b_\ell(K+\tilde{k},Q+\tilde{q})$, where 
\begin{equation}\label{eq:Cond_trans}
\phi_\ell(K+\tilde{k},Q+\tilde{q})=\left\langle{b_\ell(K+\tilde{k},Q+\tilde{q})}\right\rangle=\phi_\ell(K,Q) \delta_{\tilde{k},0} \delta_{\tilde{q},0},
\end{equation}
is the condensate order parameter in momentum space. In the second equality of Eq.\ (\ref{eq:Cond_trans}), we have assumed that the clusters are large enough, such that the minima of the non-interacting dispersion can be reproduced by the quasi-momenta of the $C_x\times C_y$ clusters (i.e.\ $\mbf{K}_{n,m}=\left(2n\pi/C_x,2m\pi/C_y\right)$). In this case, for local translational-invariant interactions, we can safely assume to have condensation only in the cluster momenta (i.e.\ at $\tilde{k}=\tilde{q}=0$).

Neglecting the terms $\mathcal{O}(\delta b^\dagger \delta b)$ in the inter-cluster couplings, we find the effective cluster-local RCMF Hamiltonian,
\begin{eqnarray}\label{eq:Ham-eff}
H^{\rm{eff}}  = &-\sum_{\mbf{R},\mbf{R}'} \left[\overline{t}_{\mbf{R},\mbf{R}'} b_{\mbf{R}}^\dagger b_{\mbf{R}'}+ \delta{t}_{\mbf{R},\mbf{R}'} (\phi_{\mbf{R}}^* b_{\mbf{R}'}+ b_{\mbf{R}}^\dagger \phi_{\mbf{R}'}) +\delta{t}_{\mbf{R},\mbf{R}'} \phi_{\mbf{R}}^* \phi_{\mbf{R}'}\right] +\nonumber\\
&+ \sum_{\mbf{R}} \left[\frac{U}{2} n_{\mbf{R}}\left(n_{\mbf{R}} -1\right)-\mu n_{\mbf{R}}\right] ,
\end{eqnarray}
where $\mbf{R}=(X,Y)$ with $X\in\left[0,C_x-1\right]$ and  $Y\in\left[0,C_y-1\right]$. Here, $\overline{t}_{\mbf{R},\mbf{R}'}$ is the hopping-amplitude of the coarse-grained dispersion from Eq.\ (\ref{eq:cluster-disp}), while the factor $\delta{t}_{\mbf{R},\mbf{R}'}$ arises from $\delta \varepsilon_{\ell,\ell'}(K,Q)$ and can be computed as
\begin{equation}\label{eq:t-relation}
\delta{t}_{\mbf{R},\mbf{R}'} = t_{\mbf{R},\mbf{R}'} - \overline{t}_{\mbf{R},\mbf{R}'} .
\end{equation}
 For details on the hopping amplitudes and their values for different lattices see \ref{sec:Ly8}. 
 
The stationary solutions with respect to the condensate $\phi_{\mbf{R}}$ can be found by requiring $\frac{\delta \Omega^{\rm{eff}}}{\delta \phi_{\mbf{R}}}=0$, where $\Omega^{\rm{eff}}$ is the free energy of the Hamiltonian (\ref{eq:Ham-eff}). This yields the self-consistency condition
\begin{equation}\label{eq:self-con}
\phi_{\mbf{R}}= \langle{b_{\mbf{R}}}\rangle,
\end{equation}
which can be solved iteratively. For a more detailed derivation of the method and technical details, see Ref.\ \cite{Hugel17}.

It should be noted that, as the approximation neglects quadratic fluctuations between particles on neighboring clusters (which we know matter at this level of accuracy, cf.~\cite{BDMFT1,BDMFT}), RCMF tends to systematically overestimate the condensate order parameter at the expense of gapped phases. This can probably be improved, but is left for future work. Note, however, that RCMF reduces to ED  (up to a renormalization of the parameters originating from coarse graining in momentum space) where it finds gapped phases: It is thus well equipped to find non-trivial phases and only approaches the thermodynamic limit in a non-conventional but mathematically sound way. Since conventional ED lacks $U(1)$-symmetry-breaking, regions of the phase diagram where both methods agree are therefore a strong indication of trustworthiness.

\section{Results}\label{sec:results}

\begin{figure}
 \centering
\includegraphics[width=300pt]{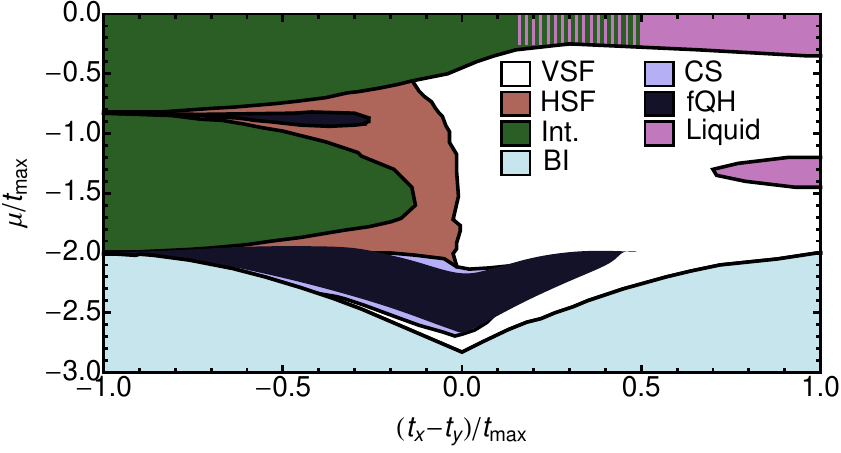}
\caption{\label{fig:phasediag} Ground-state phase diagram of the HHMm on the $L_y=4$ cylinder as computed with RCMF as a function of chemical potential $\mu/t_{\rm max}$ and hopping anisotropy $\left(t_x-t_y\right)/t_{\rm max}$. The observed phases are: band insulator (BI, light blue), gapped  phases at integer fillings $\nu=1, 2$ (Int., green), quasi-one-dimensional analogues of fractional quantum Hall phases at fillings $\nu=1/2, 3/2$ (fQH, dark grey), gapless liquid (Liquid, pink), and superfluid phases with different patterns: vertically striped (VSF, white), horizontally striped (HSF, brown) or with checkerboard order (CS, dark blue). The phases in the purely one-dimensional case are an uncondensed superfluid at $\left(t_x-t_y\right)/t_{\rm max}=1$, and gapped decoupled $4$-site rings at $\left(t_x-t_y\right)/t_{\rm max}=-1$  (except for the band-insulator at density $n=0$). In the dashed region around zero chemical potential the results are inconclusive whether the phase is gapped (Int.) or gapless (Liquid). The fractional plateau at $\nu=1/2$ was computed using an $8\times 4$ cluster, unlike the results for higher chemical potentials/densities which were computed using a $4\times 4$ cluster, see also   \ref{sec:Scaling}. The plateau at $\nu=3/2$ is most likely larger than the one found here (see Fig\ \ref{fig:spgaps}c), however both RCMF and ED are inconclusive on the exact phase boundaries within the accessible cluster-/system-sizes.}
\end{figure}

The ground-state phase diagram of the quasi-one-dimensional model computed with RCMF (see Sec.~\ref{sec:rcmf}) is shown in Fig.~\ref{fig:phasediag} in terms of the hopping anisotropy $\left(t_x-t_y\right)/t_{\rm max}$ and the chemical potential $\mu/t_{\rm max}$. In the limit $\left(t_x-t_y\right)/t_{\rm max}=1$ (i.e.\ $t_y=0$), the system consists of four decoupled infinite chains each exhibiting an uncondensed one-dimensional superfluid (or band-insulating) ground-state. At $\left(t_x-t_y\right)/t_{\rm max}=-1$ (i.e.\ $t_x=0$) instead, the cylinder reduces to an infinite set of decoupled $4$-site rings which -- depending on the filling -- can exhibit gapped phases. The phase diagram therefore strongly differs at negative anisotropies from the one observed in the fully two-dimensional model \cite{Hugel17}. The fact that the phases discussed in the following are in part entirely related to the quasi-one-dimensional geometry of the lattice is further evidenced by the fact that the situation changes drastically as soon as $L_y$ is changed from $4$ to $8$ with two unit cells in the $y$-direction, as discussed in  ~\ref{sec:Ly8}, where the $L_y=8$ results are much closer to the fully two-dimensional results than the ones for $L_y=4$.

\begin{figure}
 \centering
\includegraphics[width=300pt]{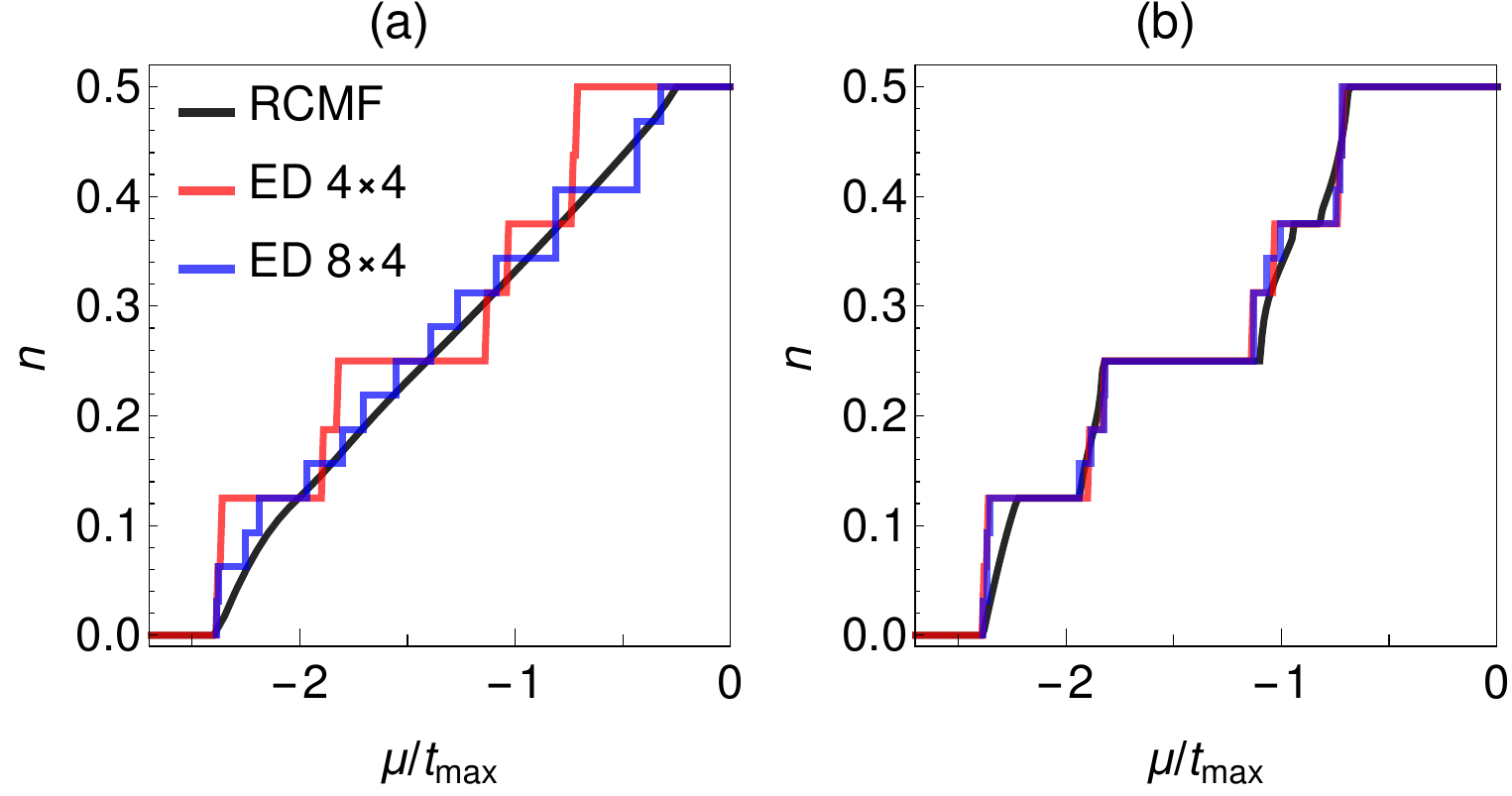}
\caption{\label{fig:den-en} Comparison between results computed with RCMF (black) and ED on system sizes $4\times 4$ (red) and $8\times 4$ (blue) with periodic boundary conditions. Particle density $n$ as a function of chemical potential $\mu/t_{\rm max}$ at hopping anisotropy $\left(t_x-t_y\right)/t_{\rm max}=0.35$ (a) and $\left(t_x-t_y\right)/t_{\rm max}=-0.35$ (b).}
\end{figure}

Before discussing the different phases in more detail, let us illustrate the role of sign asymmetry in the hopping anisotropy as well as perform a benchmarking of RCMF against ED by studying the density as a function of chemical potential, shown in Fig.~\ref{fig:den-en}. If $\mu$ is sufficiently negative the model is in a trivial band-insulating (BI) phase with zero particles for either anisotropy.
For hopping anisotropy  $\left(t_x-t_y\right)/t_{\rm max}=0.35$ (Fig.~\ref{fig:den-en}a), RCMF always finds a condensed ground-state except in a narrow region at density $n=1/2$. The ED results (which can only exhibit plateaus at densities commensurable with the system size and are computed by comparing the grand-canonical energies of the respective particle-number sectors) tend towards the continuous RCMF results as $L_x$ is increased, as all observed plateaus quickly shrink with system size (again excluding the narrow region at density $n=1/2$, which remains gapped in either method). For hopping anisotropy $\left(t_x-t_y\right)/t_{\rm max}=-0.35$ (Fig.~\ref{fig:den-en}b),  the agreement between RCMF and ED increases further, as the RCMF ground-state now also shows plateaus at integer ($\nu=1,2$, i.e.\ $n=1/4,1/2$) and fractional ($\nu=1/2,3/2$, i.e.\ $n=1/8,3/8$) fillings. The fractional plateaus, which look almost system size independent in ED, are however smaller in RCMF. In the limit of low densities however, we are able to extend the RCMF cluster from a size of $4\times 4$ to $8\times 4$ (see   \ref{sec:Scaling}). As can be seen in Figs.\ \ref{fig:spgaps}a and \ref{fig:rcmfcomparison} for the plateau at $\nu=1/2$ this increases the accuracy substantially, finding excellent agreement with ED.

\subsection{Symmetry-broken phases}\label{sec:condensed}

The ground-state away from integer ($\nu=1,2$) and half-integer ($\nu=1/2, 3/2$) filling is always condensed, exhibiting different density and condensate modulations as a function of chemical potential and hopping anisotropy. The three resulting phases we will discuss in the following exhibit first order phase transitions.

\begin{figure}
\centering
\includegraphics[width=300pt]{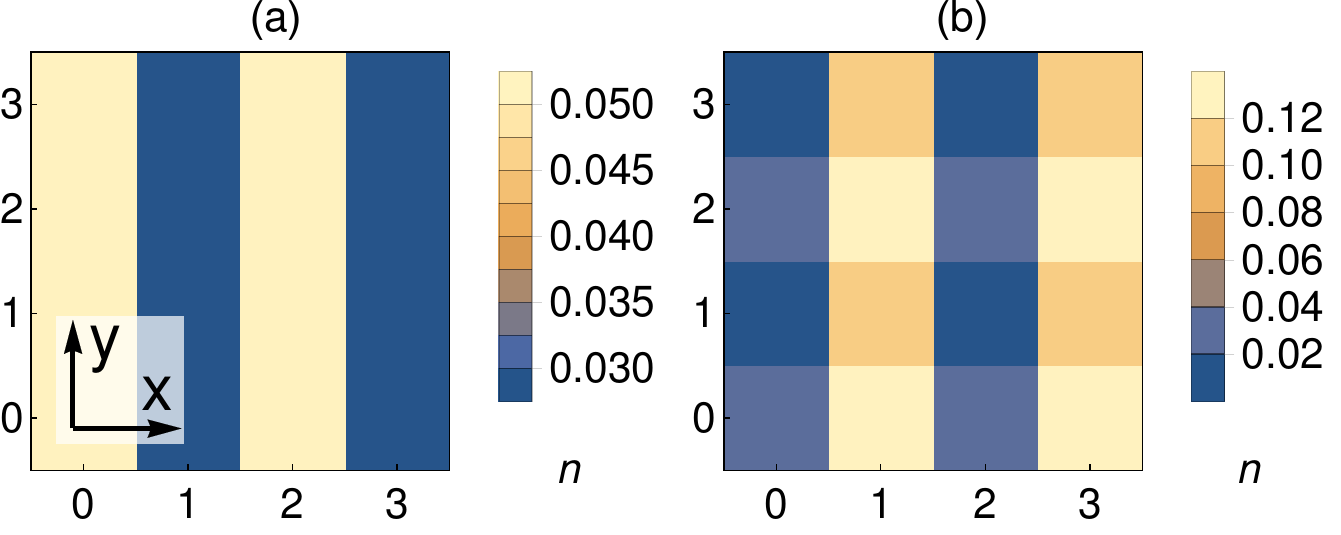}
\caption{\label{fig:snapshot} Patterns of the particle density $n$ in two different superfluid ground-state phases: (a) The vertically striped superfluid (VSF) at $\left(t_x-t_y\right)/t_{\rm max}=0.35$ and  $\mu/t_{\rm max}=-2.3$; (b) The checkerboard superfluid (CS) at $\left(t_x-t_y\right)/t_{\rm max}=-0.35$ and  $\mu/t_{\rm max}=-2.3$.}
\end{figure}

At positive anisotropy a large part of the phase diagram is occupied by the vertically striped superfluid (VSF) where both the total and the condensate density exhibit vertical stripes (see Fig.~\ref{fig:snapshot}a). With increasing negative anisotropy, these patterns are rotated by $\pi/2$ and the superfluid becomes horizontally striped (HSF). Finally, in the vicinity of filling $\nu=1/2$ the ground-state exhibits a checkerboard  superfluid pattern, apparently due to doping mechanisms on top of the degenerate ground-state (CS, see Fig.~\ref{fig:snapshot}b). 

\subsection{\texorpdfstring{$U(1)$-symmetry preserving phases}{U(1)-symmetry preserving phases}}\label{sec:uncondensed}

In the absence of a finite condensate order parameter $\phi_{\mbf{R}}$, the Hamiltonian (\ref{eq:Ham-eff}) reduces to a finite $U(1)$-symmetry-preserving $4\times 4$ torus with periodic boundaries. We can therefore turn to ED in order to further analyze the properties of $U(1)$-symmetry preserving phases (the only difference is a rescaling of the energy unit). 

In order to extrapolate to the thermodynamic limit $L_x\rightarrow\infty$ we apply the twisted boundary conditions $\Psi(x+L_x)=e^{i\theta_x}\Psi(x)$. This allows us to estimate the many-body gap as
\begin{equation}\label{eq:gap}
\Delta E = \min_{\theta_x} (\varepsilon_1(\theta_x) - \varepsilon_{\rm GS}(\theta_x)),
\end{equation}
where $ \varepsilon_{\rm GS}$ and $\varepsilon_1$ are the energies of the ground-state(s) and the first excited state, respectively. In addition, by analyzing the behavior of the quasi-one-dimensional $h$-vector \cite{Hugel17} as a function of the twisting angle $\theta_x$ we can extrapolate the topological properties of the system. If the $h$-vector shows a closed-loop as a function of $\theta_x$, and the ground-state does not mix with the excited states such that (\ref{eq:gap}) stays finite, it implies that the many-body ground-state is adiabatically translated by a single site in $y$-direction during one charge-pump cycle, resulting in different quantized Hall conductances depending on the filling. For details on twisted boundaries and the $h$-vector, see  ~\ref{sec:twisted}.

\subsubsection{Integer filing}\label{sec:uncondensed-int}

\begin{figure}
\centering
\includegraphics[width=300pt]{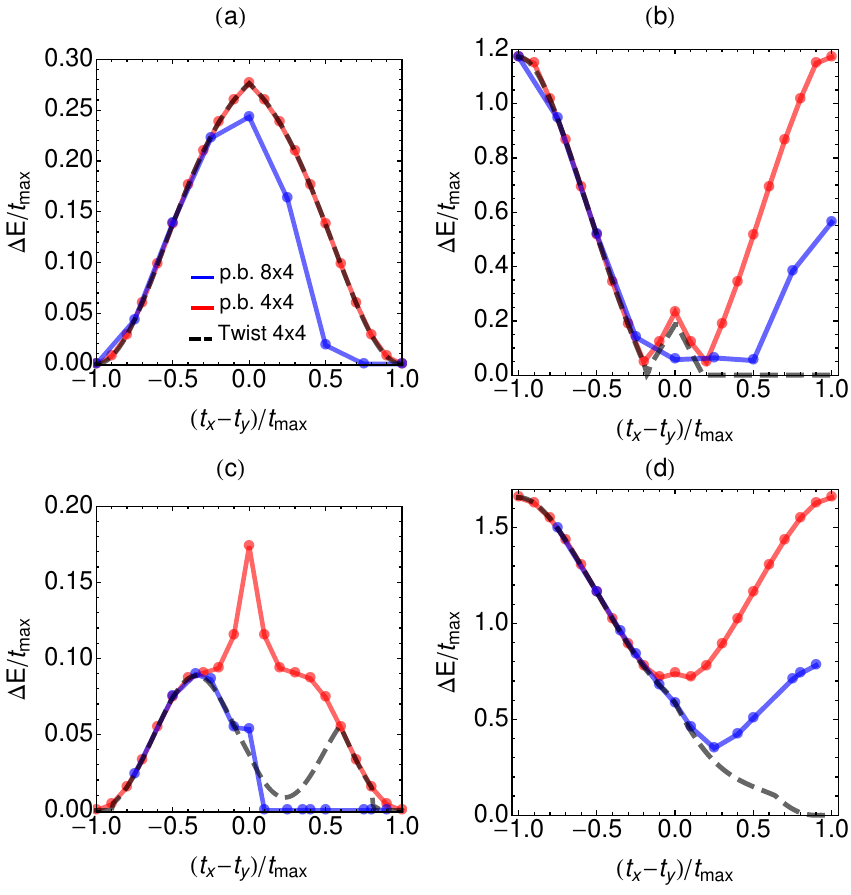}
\caption{\label{fig:mbgaps} Many-body gaps $\Delta E/t_{\rm max}$ as a function of hopping anisotropy $\left(t_x-t_y\right)/t_{\rm max}$ computed using twisted boundaries on a $4\times 4$ system (black dashed), periodic boundaries on a $4\times 4$ system (red), and periodic boundaries on a $8\times 4$ system (blue) for fillings $\nu=1/2$ (a), $\nu=1$ (b), $\nu=3/2$ (c), and $\nu=2$ (d).}
\end{figure}

\begin{figure}
\centering
\includegraphics[width=300pt]{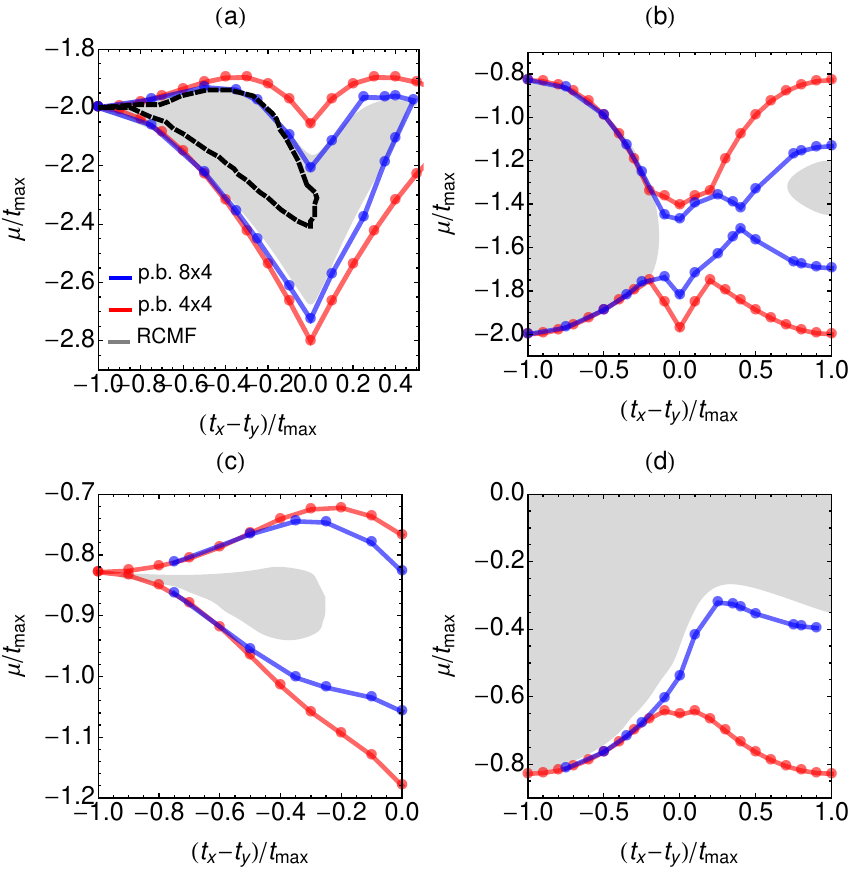}
\caption{\label{fig:spgaps} Regions of the ground-state phase diagram computed with RCMF where we observe non-trivial phases with $\phi=0$ (grey), compared with the single particle gaps measured with ED using periodic boundaries on system sizes $4\times 4$ (red) and $8\times 4$ (blue) for fillings $\nu=1/2$ (a), $\nu=1$ (b), $\nu=3/2$ (c), and $\nu=2$ (d). In panel (a) the RCMF plateau is computed using an $8\times 4$ cluster, while the RCMF results using a  $4\times 4$ cluster (which is employed for all other fillings) are shown as a black dashed line.}
\end{figure}

At (low enough) negative hopping anisotropy we observe non-degenerate ground-states at integer fillings $\nu=1$, and $2$ (see ``Int.''\ in Fig.~\ref{fig:phasediag}). In Figs.~\ref{fig:mbgaps}b and \ref{fig:mbgaps}d we show the many-body gap computed with ED for these fillings as a function of hopping anisotropy. The gap computed with (\ref{eq:gap}) on a $4\times 4$ system, as well as the scaling when going from $L_x=4$ to $L_x=8$ clearly indicate a gapped ground-state. Furthermore, as shown in Figs.~\ref{fig:spgaps}b and \ref{fig:spgaps}d, the single-particle gap computed with ED agrees perfectly with the region where we observe the gapped integer filling phases in RCMF.

For $\nu=2$ a known candidate for gapped non-trivial many-body states in the HHMm is the bosonic integer quantum Hall (biQH) phase with transverse conductance $\sigma_{xy}=2$ \cite{Kol_91,Senthil,Mol_C2,Sterdy_PRB_15,He_HHM,Andrews_arxiv}. While this phase has been found in the HHMm with hard-core bosons at lower fluxes \cite{He_HHM}, the case of $\Phi=\pi/2$ is special, as filling $\nu=2$ corresponds to a density of $n=1/2$, where hard-core bosons show a $\mathcal{C}\mathcal{T}$-symmetry (as discussed in Sec.\ \ref{sec:model}). We checked numerically that this symmetry is not spontaneously broken, by computing the overlap of the ED ground-state with its $\mathcal{C}\mathcal{T}$-transform (i.e.\ the complex-conjugated ground-state after a particle-hole transform), yielding always $1$. The Hall conductance of this ground-state can therefore only be $\sigma_{xy}=0$.

At filling $\nu=1$, a biQH phase can only be expected in the presence of two different bosonic species \cite{iQH_11} (i.e.\ filling $\nu=1+1$). Another candidate for gapped phases is the non-abelian Moore-Read state \cite{Moore_Read,Ster_Read}, which however is characterized by a degenerate ground-state, not observed here. 

As discussed in  ~\ref{sec:twisted}, the $h$-vector winds once around the origin as a function of the twisting angle for both fillings. For $\nu=2$ this implies the quantized transverse transport of a single particle-hole pair during one charge-pumping cycle, resulting in a total Hall conductance of $\sigma_{xy}=0$, consistent with the charge conjugation-symmetry of (\ref{eq:ct}). For $\nu=1$, on the other hand, this implies the transport of a single particle, and thereby a Hall conductance of $\sigma_{xy}=1$, as would be observed in a fermionic integer quantum Hall effect \cite{Klitzing_QH}.

Especially the latter phase may seem surprising, as such a bosonic phase with odd Hall conductance is expected to be either gapless, or show intrinsic topological order and thereby a degenerate ground-state \cite{Senthil,Lu_17,Moore_Read,Ster_Read}. However, while in the two-dimensional lattice phases with $\phi=0$ at these fillings are found to be always gapless liquids \cite{Hugel17}, in the quasi-one-dimensional setup these are connected to the limit of decoupled $4$-site rings at $t_x=0$, where hard-core bosons behave as free fermions gapped by the finite size of the rings, providing an intuitive explanation for the ``fermionic'' behavior of the $\nu=1$ phase. In fact, the liquid phase of the two-dimensional model \cite{Hugel17} -- just as metallic Fermi-liquid-like phases of hard-core bosons predicted in the lowest Landau level \cite{Zeng_16,Read,Sheng_Dis}  -- shows an average Hall response of $1$ which is however not protected against disorder due to its gapless nature. Here, it appears that this phase is gapped through the finite-size in $y$-direction.

Eventually, when going to positive hopping anisotropies, the gaps computed with (\ref{eq:gap}) and the strongly size-dependent scaling of the gaps with periodic boundaries shown in Figs.~\ref{fig:mbgaps}b and \ref{fig:mbgaps}d indicate a gapless ground-state for both fillings. These phases are equivalent to the anisotropic gapless liquid observed in the fully two-dimensional model \cite{Hugel17}. In fact, at positive anisotropies, those phases are connected to the limit of decoupled infinite chains at $t_y=0$, where the hard-core bosons are in a superfluid ground-state.

\subsubsection{Fractional filling}\label{sec:uncondensed-frac}

In the region of negative hopping anisotropies we observe commensurate phases with at fillings $\nu=1/2$, and $3/2$. For both fillings the system is characterized by a two-fold degenerate ground-state, and is gapped for negative (and low positive) anisotropies, as evidenced by the ED results shown in Figs.~\ref{fig:mbgaps}a and \ref{fig:mbgaps}c. 

As for the integer filling phases, the $h$-vector shows a closed loop as a function of the twisting angle (see  ~\ref{sec:twisted}) indicating Hall conductances of $\sigma_{xy}=1/2$, and $3/2$, respectively. In the two-dimensional case a fQH phase at filling $\nu=1/2$ has already been observed in previous ED \cite{ED_05,Mol_CFM,ED_07}, variational Gutzwiller mean field \cite{Onur2}, and DMRG \cite{He_HHM,dmrg_hhm2} studies, while the one at $\nu=3/2$ has not been observed in the two-dimensional limit. As for the integer phases, which are only gapped on the cylinder, the fractional phases therefore appear to be much more stable on a quasi-one-dimensional geometry. The observed phases on the cylinder differ from their isotropic counterpart in the two dimensional continuum by a weak charge density wave order (see \ref{sec:twisted}). However, while for the integer filling phases the many-body gap closes for twisting in the $y$-direction and $\sigma_{yx}\neq\sigma_{xy}$ is not quantized, the fractional phases stay gapped for twisting in both directions, and therefore show a truly two-dimensional Hall response $\sigma_{yx}=\sigma_{xy}$. 
We conclude that these phases are quasi-one-dimensional analogues of fQH phases, which continuously develop into their two-dimensional counterparts as the circumference of the cylinder is increased \cite{TT1,TT2,TT3,Grusdt_Lad}. A similar case has been discussed in the $1+1$-dimensional two-leg ladder with an additional external parameter in Ref.\ \cite{Grusdt_Lad}.

While showing the same general trend, the single particle gaps predicted by ED for the fQH phases in Figs.~\ref{fig:spgaps}a and \ref{fig:spgaps}c are significantly larger than the ones predicted by RCMF using a $4\times 4$ cluster, which tends to underestimate long-range-entangled phases. However, when employing an $8\times 4$ cluster for low densities (see   \ref{sec:Scaling}), the results on the $\nu=1/2$ plateau show excellent agreement with ED. Unfortunately, for the $\nu=3/2$ plateau such a cluster size is out of reach for RCMF, and also ED appears to not be converged with system size, such that we can only conclude that the phase boundaries of the fractional phase at $\nu=3/2$ lie in between the ones found with ED and RCMF.
 
 In the isotropic two-dimensional system, the fractional phase at filling $\nu=1/2$ predicted by other methods \cite{ED_05,Mol_CFM,ED_07,He_HHM,dmrg_hhm2,Onur2} is found to be slightly metastable in RCMF~\cite{Hugel17}. However, seeing the scaling when increasing $C_x$ to $8$, we expect that an $8\times 8$ cluster would be needed to fully capture the ground-state behavior at $\nu=1/2$ in the two-dimensional case. Finally, we note that both our RCMF results and our ED results do not point to a gapped phase at filling $\nu=2/3$ claimed in Ref.\ \cite{He_HHM} (see also \ref{sec:nu_23}).

\section{Conclusion}\label{sec:conc}

In this work we studied the ground-state properties of the Harper-Hofstadter-Mott model with quarter-flux per plaquette and hard-core bosons on a quasi-one-dimensional lattice consisting of a single flux quantum along the $y$-direction. 

We found that the stability of topologically non-trivial phases is significantly enhanced by the quasi-one-dimensional geometry. The ground-state phase diagram features quasi-one-dimensional analogues of fractional quantum Hall phases at fillings $\nu=1/2$ and $\nu=3/2$, where the latter was not observed in the two-dimensional system. We further observed new gapped non-degenerate ground-states at fillings $\nu=1$ (characterized by an odd ``fermionic'' Hall conductance of $\sigma_{xy}=1$) and at filling $\nu=2$ -- with total zero Hall conductance, but characterized by the quantized transverse transport of a single particle-hole pair as Hall response. 

By systematically comparing RCMF and ED, which approach the thermodynamical limit from opposite sides (the first method favours gapless, the second one gapped phases), we are able to give conclusive quantitative answers on the phase boundaries of gapped phases.

These unconventional integer filling phases -- which do not exist in the two-dimensional case -- illustrate the peculiarity of quasi-one-dimensional geometries in topological systems. The increased stability of the gapped phases in this setup -- which could be realized by mapping the finite spatial direction $y$ onto internal degrees of freedom -- could facilitate the experimental search for topologically non-trivial bosonic phases.

\section*{Acknowledgments}
The authors would like to thank A. Hayward, A. M. L\"auchli, G. M\"oller, F. Pollmann, and H. U. R. Strand for fruitful discussions. FK is supported by the Likar foundation. DH and LP are supported  by  FP7/ERC  Starting  Grant  No.\ 306897 and acknowledge funding by DFG through NIM-2.

\appendix
\section{Analysis of gapped phases with exact diagonalization}\label{sec:twisted}

By examining the spectrum on a finite system, twisted boundaries offer a reasonable tool to compensate for finite-size effects in ED calculations regarding the robustness of spectral gaps. In our calculations on the $4\times 4$ torus, the $y$-direction is treated exactly (as $L_y=4$). We can therefore only introduce a twisted boundary angle in the $x$-direction, $\theta_x$, defined as
\begin{equation}
\Psi(x+L_x)=e^{i\theta_x}\Psi(x).
\end{equation}

\begin{figure}
\centering
\includegraphics[width=300pt]{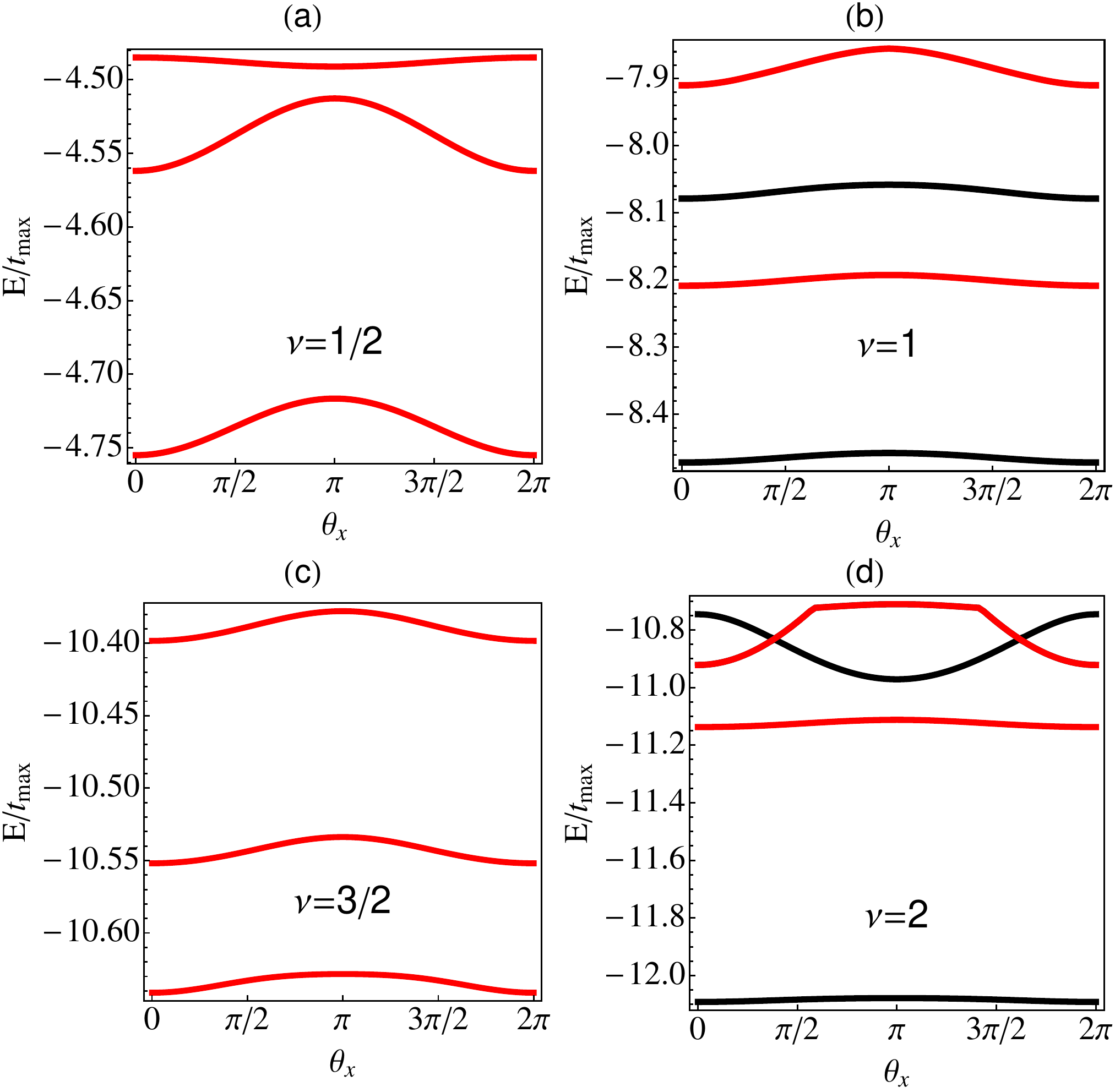}
\caption{\label{fig:twisting-spectrum-minus} Lowest $6$ ED eigenvalues of a $4\times 4$ system as a function of the twisted boundary angle $\theta_x$ for hopping anisotropy $\left(t_x-t_y\right)/t_{\rm max}=-0.35$, and fillings $\nu=1/2$ (a), $\nu=1$ (b), $\nu=3/2$ (c), and $\nu=2$ (d). Non-degenerate eigenstates are shown in black, doubly-degenerate ones in red. }
\end{figure}

\begin{figure}
\centering
\includegraphics[width=300pt]{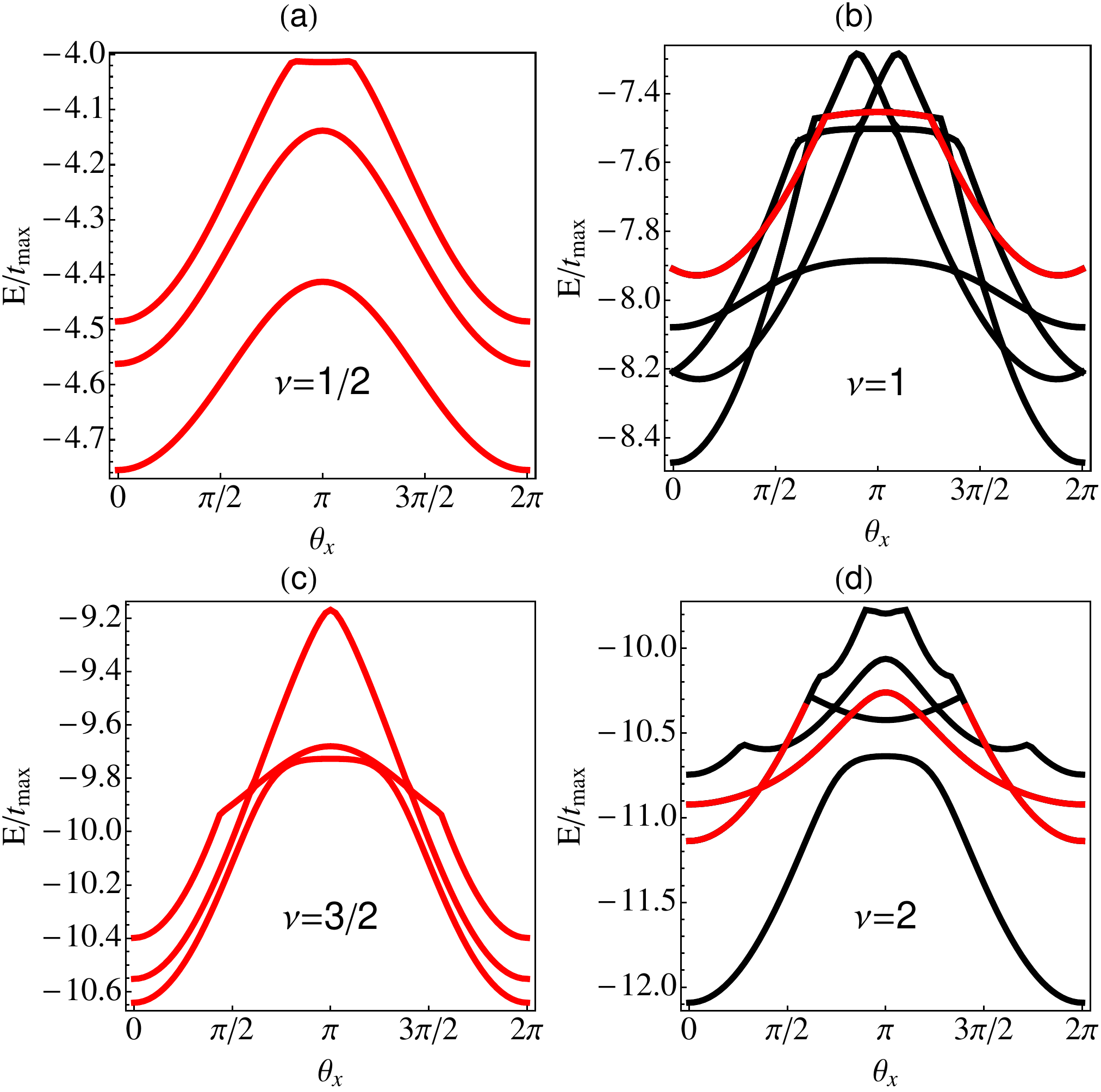}
\caption{\label{fig:twisting-spectrum-plus} Lowest $6$ ED eigenvalues of a $4\times 4$ system as a function of the twisted boundary angle $\theta_x$ for hopping anisotropy $\left(t_x-t_y\right)/t_{\rm max}=0.35$, and fillings $\nu=1/2$ (a), $\nu=1$ (b), $\nu=3/2$ (c), and $\nu=2$ (d). Non-degenerate eigenstates are shown in black, doubly-degenerate ones in red.}
\end{figure}

Figs.~\ref{fig:twisting-spectrum-minus} and \ref{fig:twisting-spectrum-plus} show the dependency of the spectrum on the twisting angle $\theta_x$ for fillings $\nu=1/2$, $1$, $3/2$ and $2$ at $\left(t_x-t_y\right)/t_{\rm max}=-0.35$ and $\left(t_x-t_y\right)/t_{\rm max}=0.35$, respectively. The six lowest eigenvalues are shown, as computed with ED on the $4\times 4$ lattice. When the spectrum mixes, the many-body ground-state on the lattice in the thermodynamic limit can be assumed to be gapless, see Eq.\ (\ref{eq:gap}). 

In Figs.~\ref{fig:twisting-spectrum-minus} and \ref{fig:twisting-spectrum-plus} we see that the ground-state at fractional fillings is $2$-fold degenerate. Furthermore, we observe how the groundstates mix with the excited ones as a function of $\theta_x$ at $\left(t_x-t_y\right)/t_{\rm max}=0.35$ and fillings $\nu=1$ and $\nu=3/2$, while the other cases shown (i.e., $\nu=1/2$ and $\nu=2$ in Fig.~\ref{fig:twisting-spectrum-plus} and all negative values of $\left(t_x-t_y\right)/t_{\rm max}=-0.35$ in Fig.~\ref{fig:twisting-spectrum-minus}) appear to be gapped.

Let us now rewrite  the Hamiltonian (\ref{eq:Ham-diag}) into the form used in Ref.\ \cite{Hugel17} suited to introduce the $\langle{\hat{h}}\rangle$-vector. To this end, we define
\begin{eqnarray}\label{eq:AB}
A_y(k) &= d^\dagger_y (k) d_y (k) - d^\dagger_{y+2} (k) d_{y+2} (k) \label{eq:A} ,{,}\\
B(k) &= \frac{1}{2} \sum_{y} d^\dagger_{y+1} (k) d_y (k) + \mathrm{H.c.}, \label{eq:B}
\end{eqnarray}
the Hamiltonian (\ref{eq:Ham-diag}) now reads
\begin{equation} \label{eq:Ham-k-vh}
H = \sum_{k} \underline{v}_{k} \cdot \underline{h}_{k} ,
\end{equation}
where
\begin{equation}\label{eq:v-and-h}
\underline{v}_{k} = \begin{pmatrix}-2 t_x \cos(k) \\ -2 t_x \sin(k) \\ - 2 t_y \end{pmatrix} ,\quad \underline{h}_{k} = \begin{pmatrix}A_0(k) \\ A_1(k) \\ B(k)\end{pmatrix}.
\end{equation}

As discussed in more detail in Ref.\ \cite{Hugel17}, the expectation value $\langle{\hat{h}}\rangle=\frac{\langle{\underline{h}}\rangle}{\left|{\langle{\underline{h}}\rangle}\right|}$ can be used to measure the Hall response. The twisting angle $\theta_x$ can be seen as a magnetic flux piercing the system in $y$-direction and can be implemented by the transform
\begin{equation}
t_x\mapsto e^{i\theta_x/L_x}t_x,
\end{equation}
transforming $\underline{v}_{k}\mapsto \underline{v}_{k-\theta_x/L_x}$, or equivalently -- as long as the ground-state stays gapped -- $\langle{\hat{h}_k}\rangle\mapsto \langle{\hat{h}_{k+\theta_x/L_x}}\rangle$. 

\begin{figure}
\centering
\includegraphics[width=300pt]{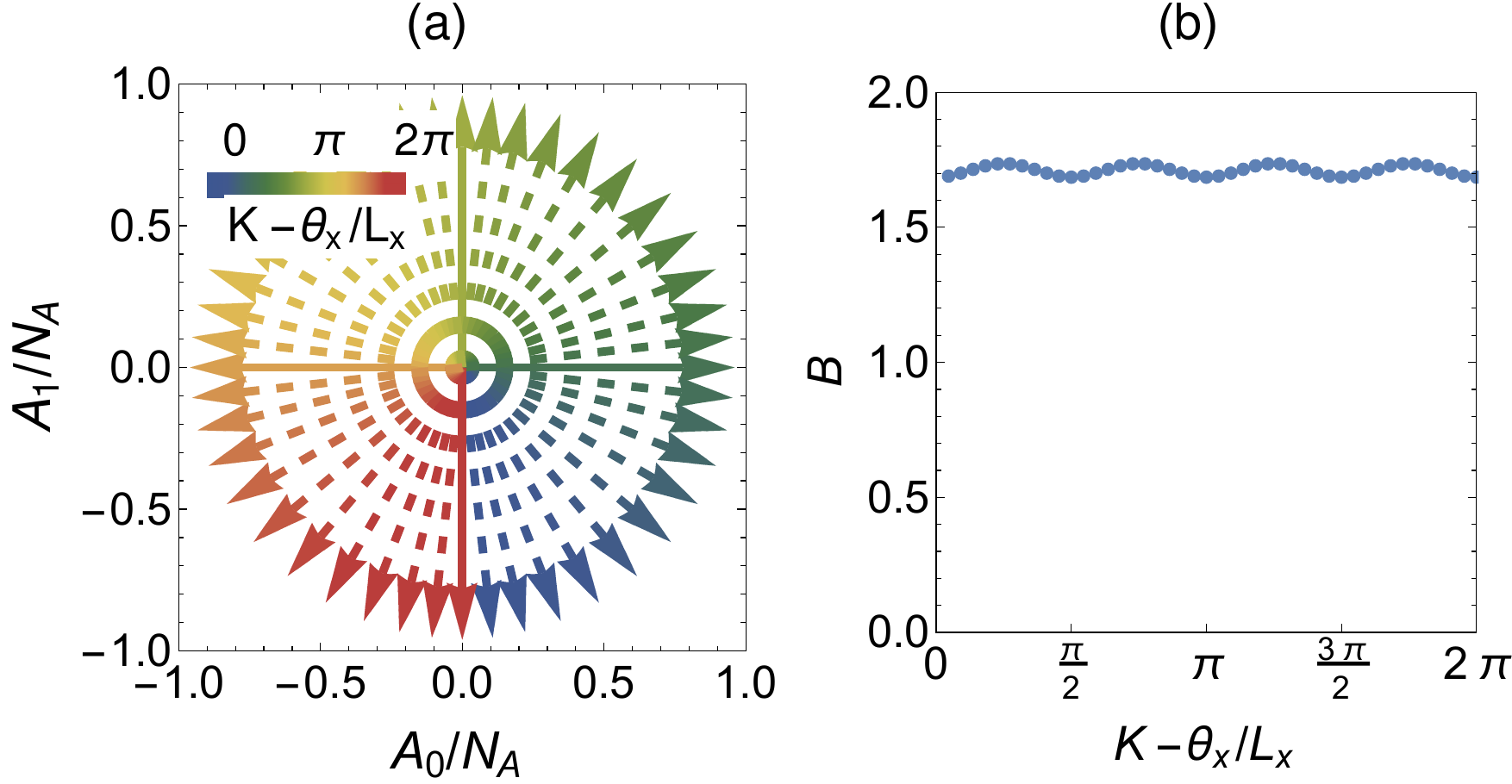}
\caption{\label{fig:winding} Winding of the $\langle{\hat{h}}\rangle$-vector as a function of momentum $K$ and twisting angle $\theta_x$ at filling $\nu=1/2$ and hopping anisotropy $\left(t_x-t_y\right)/t_{\rm max}=-0.35$. (a) Projection of $\langle{\hat{h}}\rangle$ onto the $(A_0,A_1)$-plane. The components $A_0$ and $A_1$ are normalized with $N_A=\sqrt{A_0^2 + A_1^2}$, while the coloring corresponds to different values of $K-\theta_x/L_x$. (b) $z$-component of $\langle{\hat{h}}\rangle$, i.e.\ $\langle B \rangle$, as a function of $K-\theta_x/L_x$.}
\end{figure}

As shown in Fig.~\ref{fig:winding}, this causes the rotation of $\langle{\hat{h}}\rangle$ in response to the twisting angle in the $\left(A_0,A_1\right)$-plane. During $L_x/4$ subsequent charge-pumping processes (i.e.\ $\theta_x:0\rightarrow 2\pi \cdot L_x/4$) the many-body ground-state transforms as $A_0\mapsto A_1$ and $A_1\mapsto -A_0$, and therefore $\langle d^\dagger_y (k) d_y (k)\rangle\mapsto \langle d^\dagger_{y+1} (k) d_{y+1} (k)\rangle$. This means that the many-body ground-state is adiabatically translated by a single site in $y$-direction. For a single charge-pump process ($\theta_x:0\rightarrow 2\pi$) and filling $\nu$ (i.e.\ total number of particles $N=\nu L_x$ for $L_y=4$) this amounts to the transverse transport of $N=\nu 4$ particles by one site, or equivalently $\nu$ particles through the full system (i.e.\ by $L_y=4$ sites) in $y$-direction.

The winding shown in Fig.~\ref{fig:winding} for $\nu=1/2$ therefore implies a Hall conductance of $\sigma_{xy}=1/2$. Similarly, $\nu=1$ and $\nu=3/2$ correspond to $\sigma_{xy}=1$, and $\sigma_{xy}=3/2$, respectively. The winding in the $\nu=2$ case is special, due to the charge-conjugation symmetry at $n=1/2$ [see Eq.\ (\ref{eq:ct})]. For each particle being transported there is also a hole being transported, resulting in a total quantized transport of a particle-hole pair, and zero Hall conductivity.

Unlike at fractional fillings, for the integer filling phases -- as their many-body gap is a direct consequence of the anisotropic geometry -- the Hall conductance is highly anisotropic, in the sense that it is quantized for charge-pump processes in the $x$-direction only. As the ground-state mixes with the excited states as a function of twisting in $y$-direction (as discussed in Ref.\ \cite{Hugel17}), $\sigma_{yx}\neq\sigma_{xy}$ is not quantized, emphasizing the quasi-one-dimensional nature of these phases. It should be noted that if the $y$-direction is mapped onto four different species/internal states, such that $\langle d^\dagger_y (k) d_y (k)\rangle$ and $\langle d^\dagger_{y+1} (k) d_{y+1} (k)\rangle$ can be resolved separately in time-of-flight measurements, $\langle{\hat{h}}\rangle$ could be computed in experiment, providing a direct measurement of the Hall response in equilibrium.
\begin{figure}
\centering
\includegraphics[width=300pt]{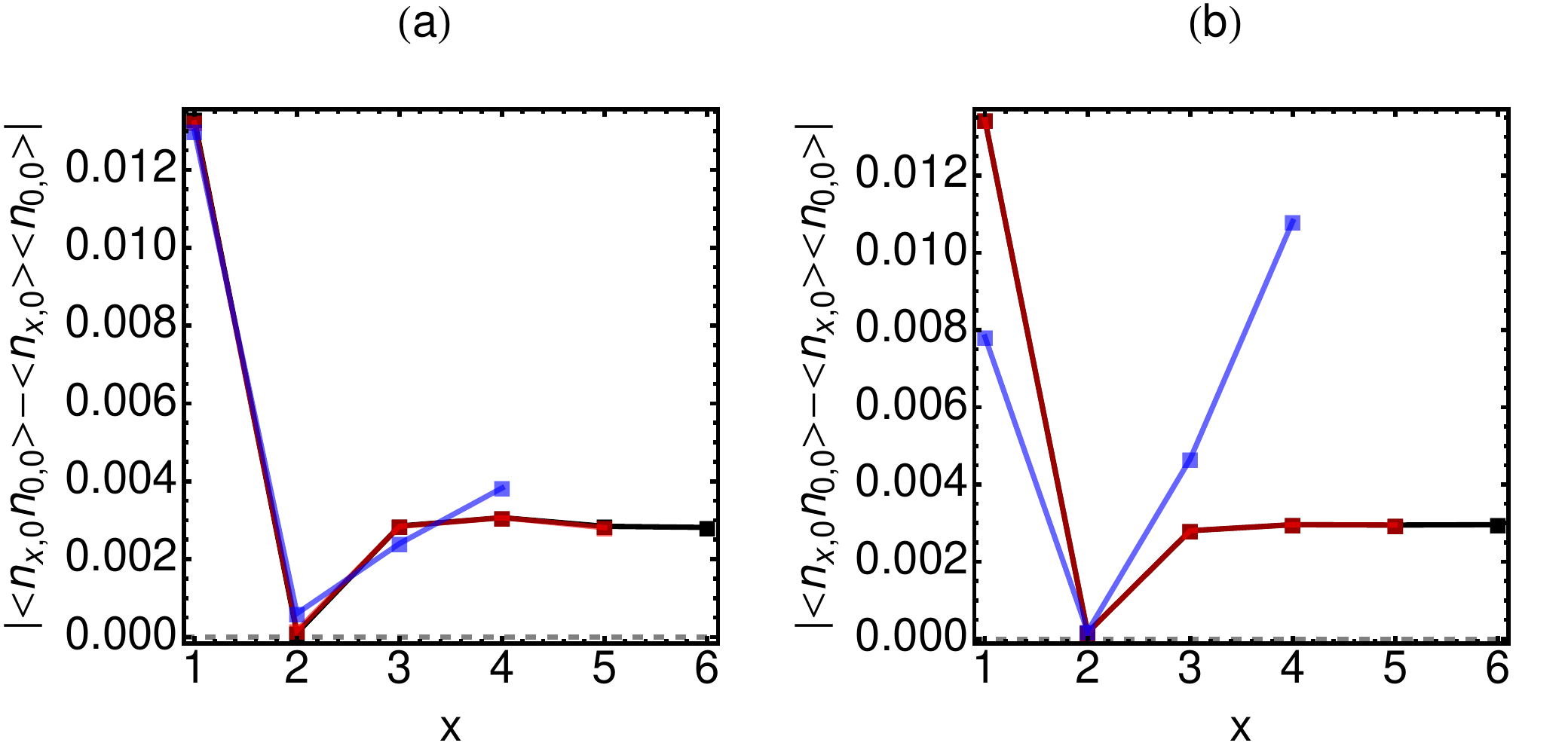}
\caption{\label{fig:cdw} Density-density correlations $\left|\langle n_{x,0}n_{0,0}\rangle-\langle n_{x,0}\rangle\langle n_{0,0}\rangle\right|$ as a function of $x$ for filling $\nu=1/2$ and $t_x=t_y$ on a cylinder with $L_y=4$ and $L_x=8$ (blue), $L_x=10$ (red), and $L_x=12$ (black). Panels (a) and (b) show the two degenerate ground-states at momentum sectors $K=0$ and $K=\pi$, respectively. Due to the periodic boundary conditions, the correlations are only shown up to the center of the cylinder at $L_x/2$.}
\end{figure}

In order to fully classify the gapped phases at fractional filling, we turn to the density-density correlations in $x$-direction $\left|\langle n_{x,y}n_{0,y}\rangle-\langle n_{x,y}\rangle\langle n_{0,y}\rangle\right|$. As shown in Fig.\ \ref{fig:cdw} for the two degenerate ground-states at filling $\nu=1/2$ and $t_x=t_y$ for $y=0$, these correlations indicate a charge density wave order: after decreasing for short distances they quickly saturate to a small finite value as a function of $x$ and appear to be converged in the system size $L_x$ for $L_x=10$ and $12$. A similar behavior is also observed for other values of $y$ and at filling $\nu=3/2$. Unlike their two-dimensional fQH counterparts in the continuum, these fractional phases are therefore characterized by a weak charge density wave order. The Hall response, on the other hand, is fully isotropic $\sigma_{yx}=\sigma_{xy}=\nu$, just as in the two-dimensional case. 

\section{Quasi-one-dimensional vs two-dimensional geometry}\label{sec:Ly8}

The quantity which differentiates between the different lattice geometries in RCMF is the coarse-grained dispersion on the $4\times 4$ cluster, i.e.\ 
\begin{equation}\label{eq:eps_bar}
\overline{\epsilon}_{\mbf{K}} = \frac{C_x C_y}{L_x L_y}\sum_{\mbf{\tilde{k}}}\epsilon_{\mbf{K}+\mbf{\tilde{k}}},
\end{equation}
where $\mbf{K}\in\left\lbrace\left(0,0\right),\left(0,\pi/2\right),\left(0,\pi\right),\left(0,3\pi/2\right)\right\rbrace$ are the quasi-momenta of the $4\times 4$ cluster (see Ref.\ \cite{Hugel17} for details).

\begin{figure}
\centering
\includegraphics[width=300pt]{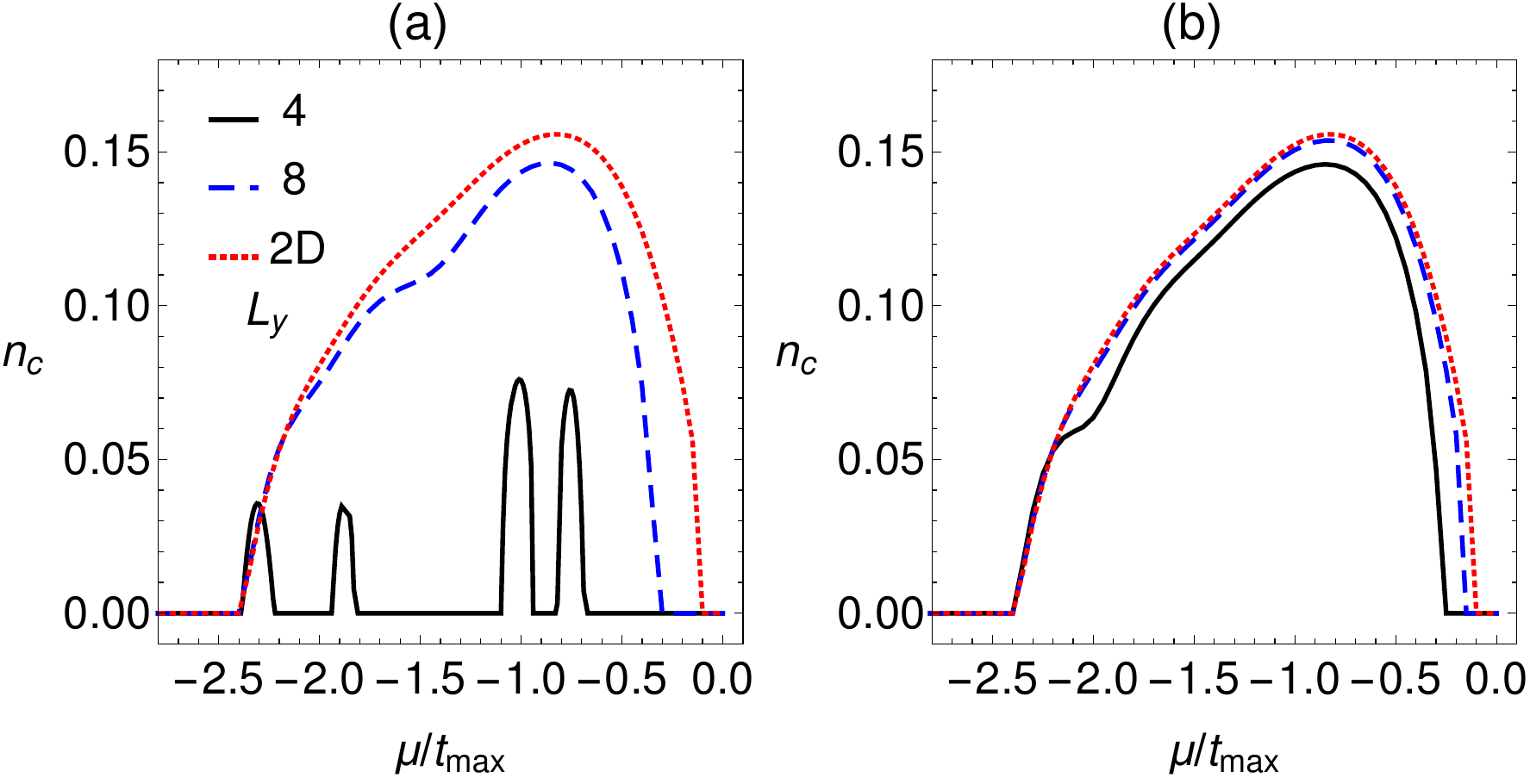}
\caption{\label{fig:condensatecomparison} Condensate density $n_c$ as a function of chemical potential $\mu/t_{\rm max}$ for different lattices, i.e.\ $L_y=4$ (solid, black), $L_y=8$ (dashed, blue) and $L_y\to\infty$ (dotted, red). In (a) the hopping anisotropy is $\left(t_x-t_y\right)/t_{\rm max}=-0.35$ and in (b) $\left(t_x-t_y\right)/t_{\rm max}=0.35$.}
\end{figure}

The general two-dimensional system can be written as $H = \sum_{\mbf{k}} H_{\mbf{k}}$, where $\mbf{k}=(k,q)$, $k$ is the quasi-momentum in $x$-direction, $q$ the quasi-momentum in $y$-direction, and
\begin{equation}\label{eq:Ham-diag-Ly8}
H_{\mbf{k}} = -\sum_{y=0}^3 \left(t_x e^{i (\frac{\pi}{2} y -k)} d_y^\dagger(\mbf{k}) d_y(\mbf{k})+t_y e^{-iq}  d_{y+1}^\dagger(\mbf{k}) d_y(\mbf{k})\right) + \mathrm{H.c.}
\end{equation}

In the thermodynamic limit in $x$-direction, assumed for all geometries ($L_x\rightarrow\infty$), $k\in\left[0,2\pi\right]$ is continuous. For $L_y=4$ considered in the main text, $q=0$, such that the coarse-grained dispersion is given by a one-dimensional integral $\overline{\epsilon}_{K}=\frac{C_x }{L_x }\int \mathrm{d} \tilde{k}\; \epsilon_{K+\tilde{k}}$, resulting in the effective hopping amplitudes in Eq.\ (\ref{eq:Ham-eff})
\begin{eqnarray}\label{eq:t-Ly4}
\overline{t}_{\mbf{R}+\mbf{e}_x,\mbf{R}} &=  \frac{2\sqrt{2}}{\pi}\; t_x\; e^{i Y \frac{\pi}{2}}, \label{eq:tx} \\
\overline{t}_{\mbf{R}+\mbf{e}_y,\mbf{R}} &=  t_y . \label{eq:ty}
\end{eqnarray}

In a system with arbitrary finite size in $y$-direction, $q=2n\pi/L_y$ with $n=0,1,\dots,L_y/4$. For a cylinder with $L_y=8$ (and therefore two flux quanta in $y$-direction) Eq.\ (\ref{eq:eps_bar}) yields the effective hopping amplitudes
\begin{eqnarray}\label{eq:t-Ly8}
\overline{t}_{\mbf{R}+\mbf{e}_x,\mbf{R}} &=  \frac{2\sqrt{2}}{\pi}\; t_x\; e^{i Y \frac{\pi}{2}}, \label{eq:t-Ly8-tx} \\
\overline{t}_{\mbf{R}+\mbf{e}_y,\mbf{R}} &=  \frac{\sqrt{2+\sqrt{2}}}{2}\; t_y . \label{eq:t-Ly8-ty}
\end{eqnarray}

Finally, in the case $L_y\rightarrow\infty$ also $q\in\left[0,\pi/2\right]$ becomes continuous, and the integral $\overline{\epsilon}_{\mbf{K}}=\frac{C_x C_y}{L_x L_y}\int \mathrm{d}\tilde{k}  \mathrm{d}\tilde{q}\; \epsilon_{\mbf{K}+\mbf{\tilde{k}}}$ yields the effective hopping amplitudes
\begin{eqnarray}\label{eq:2d}
\overline{t}_{\mbf{R}+\mbf{e}_x,\mbf{R}} &=  \frac{2\sqrt{2}}{\pi}\; t_x\; e^{i Y \frac{\pi}{2}}, \label{eq:t-2d-tx} \\
\overline{t}_{\mbf{R}+\mbf{e}_y,\mbf{R}} &=  \frac{2\sqrt{2}}{\pi}\; t_y . \label{eq:t-2d-ty}
\end{eqnarray}

In Fig.~\ref{fig:condensatecomparison} we compare the three geometries ($L_y=4$, $L_y=8$ and $L_y\rightarrow\infty$). We show the condensate density $n_c=\frac{1}{C_x C_y} \sum_{\mbf{R}} \left|{\phi_{\mbf{R}}}\right|^2$ for the three different lattices as a function of chemical potential, illustrating how the  $L_y=8$ cylinder with just two unit-cells in $y$-direction already shows much better agreement with the gapless two-dimensinal results \cite{Hugel17}, further indicating the exclusive quasi-one-dimensional nature of the $L_y=4$ cylinder.

\section{RCMF scaling at low densities}\label{sec:Scaling}

\begin{figure}
\centering
\includegraphics[width=300pt]{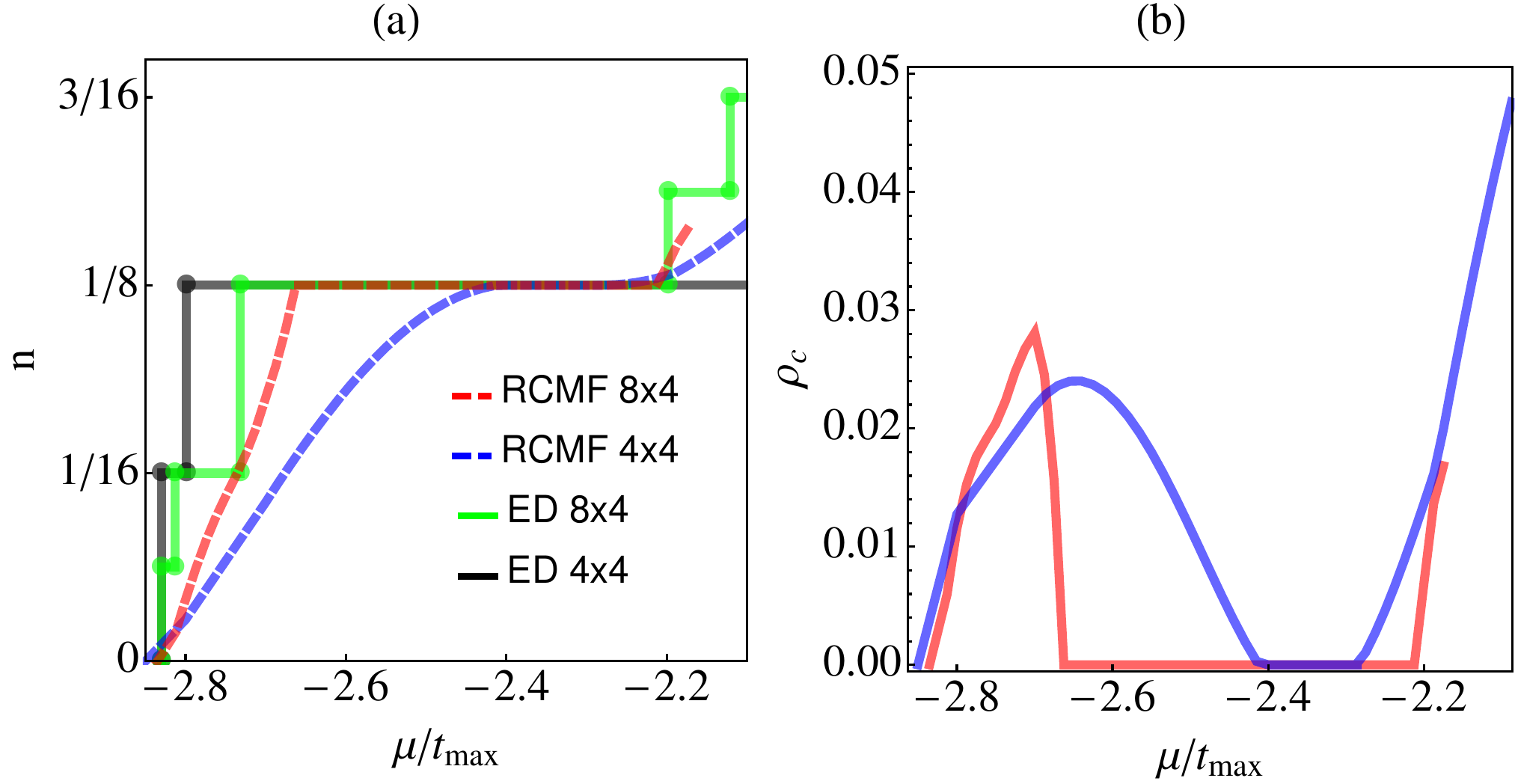}
\caption{\label{fig:rcmfcomparison} Comparison of particle density (a) and condensate density (b) as a function of chemical potential for $t_x=t_y$ computed with ED on a $4\times 4$ torus (black), ED on an $8\times 4$ torus (green), RCMF with a $4\times 4$ cluster impurity (blue) and RCMF with  an $8\times 4$ cluster impurity (red).}
\end{figure}

As mentioned in Sec.\ \ref{sec:rcmf}, for the RCMF approximation to work, it is indispensable that the momentum values of the four minima of the non-interacting dispersion at $k=0,\pi,\pm \pi/2$ can be represented within the $C_x\times 4$ cluster impurity. After the $4\times 4$ cluster employed in this work the next cluster size which is consistent with this approach would be therefore $8\times 4$, which in the absence of particle number conservation is computationally out of reach.

In the case of low density however we can restrict the basis of the $8\times 4$ cluster to the particle number sectors $N=0,1,2,\dots,6$, making the search for a symmetry-broken ground-state numerically accessible. The resulting hopping amplitudes with respect to the $4\times 4$ cluster change accordingly to

\begin{eqnarray}\label{eq:t-Lx8}
\overline{t}_{\mbf{R}+\mbf{e}_x,\mbf{R}} &=  \frac{8}{\pi}\sin\left(\frac{\pi}{8}\right)\; t_x\; e^{i Y \frac{\pi}{2}}, \label{eq:tx8x4} \\
\overline{t}_{\mbf{R}+\mbf{e}_y,\mbf{R}} &=  t_y . \label{eq:ty8x4}
\end{eqnarray}

In Fig.\ \ref{fig:rcmfcomparison} we compare results computed with such a restricted basis using an $8\times 4$ cluster in the vicinity of density $n=1/8$ (i.e.\ $\nu=1/2$), where such an approach is still controlled. We see how ED and RCMF scale differently with cluster size, converging towards each other: since ED prefers gapped phases, the fractional plateau at $\nu=1/2$ shrinks with cluster size. As RCMF prefers gapless phases, the fractional plateau increases with cluster size (see also Fig.\ \ref{fig:spgaps}a).

\section{\texorpdfstring{Filling $\nu=2/3$}{Filling nu=2/3}}\label{sec:nu_23}

Jain's sequence \cite{Jain89}, a composite fermion approach to the HHMm predicts a series of gapped phases at fillings $\nu=p/(p+1)$ for integer $p$. In this picture the fQH phase at filling $\nu=1/2$ is equivalent to the first state of Jain's sequence at $p=1$. The bosonic quantum Hall phase at $\nu=2$ with $\sigma_{xy}=2$ is equivalent to $p=-2$. Note that it does not exist for $\Phi=\pi/2$ with hard-core bosons (the charge-conjugation symmetry of Eq.\ (\ref{eq:ct}) imposes $\sigma_{xy}=0$ at $\nu=2$) but was measured for lower fluxes in Ref.\ \cite{He_HHM}.
\begin{figure}
\centering
\includegraphics[width=\columnwidth]{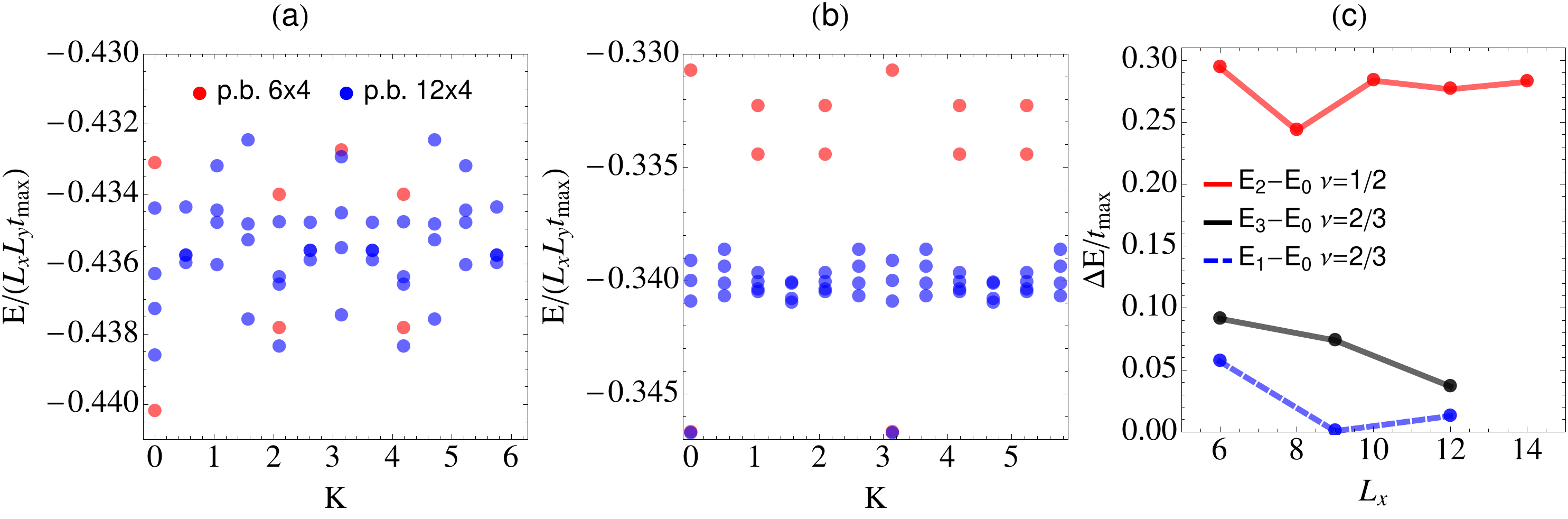}
\caption{\label{fig:nu_23} Comparison of the ED spectra at fillings $\nu=2/3$ and $\nu=1/2$. (a) and (b): Eigenstates as a function of momentum sector $K$ at $t_x=t_y$ and system sizes $6\times 4$ (red) and $12\times 4$ for fillings $\nu=2/3$ (a) and $\nu=1/2$ (b). (c) Many-body gap of the fractional phase at $\nu=1/2$ (red) and gaps at $\nu=2/3$ between the lowest and the first excited state (blue, dashed) and between the lowest and the third excited state (black) as a function of $L_x$.}
\end{figure}

Ref.\ \cite{He_HHM} further reports on a gapped phase at filling $\nu=2/3$ and no hopping anisotropy at flux $\Phi=\pi/2$, equivalent to the next state in Jain's sequence after the $\nu=1/2$ fQH (i.e.\ $p=2$), i.e.\ Halperin's $\left(211\right)$ state, which we do not observe in RCMF. We therefore turn to the ED spectrum of system sizes $L_x\times 4$ with $L_x=6$, $9$ and $12$ to look for signatures of a gapped phase at filling $\nu=2/3$ in Fig.~\ref{fig:nu_23}a. As can be seen for both $L_x=6$ and $L_x=12$ we do not observe a clear indication of a three-fold degeneracy of the ground-state expected for  Halperin's $\left(211\right)$ state.

In Fig.~\ref{fig:nu_23}c the distance between the lowest and the degenerate second and third eigenstate is shown as a function of $L_x$. While it generally decreases (and goes almost to zero at $L_x=9$), so does the gap between the third and the lowest eigenstate (which would be the many-body gap in case the ground-state were truly three-fold degenerate). For comparison, we show in Fig.~\ref{fig:nu_23}b  the spectrum for $\nu=1/2$ for the same system sizes. Here, as expected, the ground-state is two-fold degenerate, and the many-body gap to the first excited state is considerably larger (around $1/4$ of the hopping). In line with RCMF, the ED results do not seem to point toward the existence of a gapped state at filling $\nu=2/3$.

\section*{References}

\bibliography{HHHM1.bib}

\end{document}